\documentclass[sn-mathphys-num]{sn-jnl}

\usepackage{amssymb}
\usepackage{bm}
\usepackage{color}

\usepackage{amsfonts}
\usepackage{amsmath}
\usepackage{amssymb,amsthm}
\usepackage{mathtools}
\usepackage{color}

\usepackage{url}
\usepackage{hyperref}
\usepackage{graphicx}
\usepackage{booktabs}
\usepackage{multirow}
\usepackage{comment}

\numberwithin{equation}{section}             
\renewcommand{\thesection}{\arabic{section}} 

\theoremstyle{thmstyleone}%
\newtheorem{theorem}{Theorem}
\theoremstyle{thmstyletwo}%
\newtheorem{remark}{Remark}%

\theoremstyle{thmstylethree}%
\begin{document}


\title[Article Title]{Characterization of the time-dependent free Schr\"odinger operator by the Galilei invariance}




\author*[1]{\fnm{Hiromichi} \sur{Nakazato}}\email{hiromici@waseda.jp}
\affil*[1]{\orgdiv{Department of Physics}, \orgname{Waseda University}, \orgaddress{\street{Okubo 3-4-1}, \city{Shinjuku}, \postcode{169-8555}, \state{Tokyo}, \country{Japan}}}

\author[2]{\fnm{Tohru} \sur{Ozawa}}\email{txozawa@waseda.jp}
\affil[2]{\orgdiv{Department of Applied Physics}, \orgname{Waseda University}, \orgaddress{\street{Okubo 3-4-1}, \city{Shinjuku}, \postcode{169-8555}, \state{Tokyo}, \country{Japan}}}



\abstract{
The time-dependent free Schr\"odinger operator is shown to be characterized as the only linear partial differential operator of the second order that is invariant under the Galilei group in the Euclidean space-time $\mathbb R\times\mathbb R^n$. 
The method of proof depends on the analysis of the invariance of polynomials given by the application of the linear partial differential operators to monochromatic plane waves under space rotations and pure Galilei transformations.}



\maketitle

\section{Introduction}
\label{s:intro}
In this paper, we consider a general class of linear partial differential operators in the Euclidean space-time $\mathbb R\times\mathbb R^n$.  
Among others, we provide a characterization of the time-dependent free Schr\"odinger operator in terms of the Galilei transformations and local gauge transformations on scalar fields on $\mathbb R\times\mathbb R^n$.  
This means in particular that the time-dependent Schr\"odinger equation is derived from a geometric observation of the Galilei group of motion acting on scalar fields up to local gauge.  

In Classical Mechanics, the state of particles is described by the position and velocity in the configuration space $\mathbb R^n$.  
Among others, the motion of the free particles is written by space-time translations, space rotations, and pure Galilei transformations.  
In other words, the motion of the free particles is characterized by the Galilei group in $\mathbb R\times\mathbb R^n$.  

In Quantum Mechanics, the state of the particles is described on the basis of wavefunctions and particular physical quantities such as the position and momentum are given by the corresponding self-adjoint operators through ${\rm L}^2(\mathbb R^n;\mathbb C)$ scalar product.  

From the point of view of representation theory, the group of motion naturally acts on functions based on the notion of pull-back.  
In this setting, the Galilei group naturally acts on scalar fields on $\mathbb R\times\mathbb R^n$, and moreover, the notion of local gauge is introduced as a multiplication of functions $\exp(i\theta)$ with a phase function $\theta: \mathbb R\times\mathbb R^n\to\mathbb R$ \cite{Bargmann,Barut}. 
Here we follow the terminology of  ``global" and ``local" according to the convention of physics [1, 2]. 
Namely, the ``global" and ``local" gauge transformations are given by with factor $e^{i \theta}$ with $\theta \in \mathbb R$ and $e^{i \varphi}$ with $\varphi \in C^\infty (\mathbb R \times \mathbb R^n;\mathbb R)$. 
In this paper, we take this point of view and utilize the Galilei group with local gauge acting on scalar fields to study the invariance of partial differential operators.  
As is clear in the following, the procedure adopted here addressing the possible form of partial differential operators is unique and different from the previous studies where the free Schr\"odinger equation has been formally derived as the unitary irreducible representation of the Galilei group \cite{Hamermesh} (see also, e.g., \cite{LevyLeblond,Roncadelli,MusielakFry1,MusielakFry2}).

This paper is organized as follows.  
In Section \ref{s:main}, we introduce the basic notation of the Galilei group in the Euclidean space-time $\mathbb R\times\mathbb R^n$ and state our main results.  
In Section \ref{s:proof}, we prove the main results.

\section{Main Results}
\label{s:main}
In this section, we state our main results. 
For that purpose, we start introducing basic notation.  
We denote by $x=(t,\bm x)=(x^0,x^1,\ldots,x^n)$ a point in $\mathbb R\times\mathbb R^n$.  
We also use the notation
\begin{equation}
    x=\left[\begin{matrix}t\\\bm x\end{matrix}\right]=\left[\begin{matrix}x^0\\ x^1\\\vdots\\ x^n\end{matrix}\right]
\end{equation}
as a column-vector representation to which $(1+n)\times(1+n)$-matrices apply.  
Let $(\bm e_0,\bm e_1,\ldots,\bm e_n)$ be the standard basis of $\mathbb R\times\mathbb R^n$.  
Then the column vector representation of a point in $\mathbb R\times\mathbb R^n$ is given by 
\begin{equation}
    x=\left[\begin{matrix}t\\\bm x\end{matrix}\right]=x^0\bm e_0+\sum_{j=1}^nx^j\bm e_j.
\end{equation}
We equip the $n$-dimensional Euclidean space $\mathbb R^n$ with the inner product given by 
\begin{equation}
\bm x\cdot\bm y=\sum_{j=1}^nx^jy^j,
\end{equation}
which is a natural consequence of the orthonormality relations $\bm e_j\cdot\bm e_k=\delta_{jk}$, where $\delta_{jk}=0$ if $j\not=k$ and $\delta_{jj}=1$.  

Let ${\rm O}(n)$ be the orthogonal group defined by
\begin{equation}
    {\rm O}(n)=\{R\in{\rm GL}(n;\mathbb R); {}^t\!RR=R\,{}^t\!R=I\},
    \label{eq:orthogonal}
\end{equation}
where ${\rm GL}(n;\mathbb R)$ is the general linear group consisting of all invertible $n\times n$ matrices, $^t\!R$ the transposed matrix of $R$, and $I$ the identity matrix of order $n$.  
The action of $R\in{\rm O}(n)$ on $\mathbb R^n$ is understood as
\begin{equation}
    R\bm x=\left[\begin{matrix}R^1\,_1&\cdots&R^1\,_n\\
    \vdots&&\vdots\\
    R^n\,_1&\cdots&R^n\,_n\end{matrix}\right]\left[\begin{matrix} x^1\\\vdots\\ x^n\end{matrix}\right]=\Bigl(\sum_{k=1}^nR^j\,_kx^k: j\in\{1,\ldots,n\}\Bigr).
\label{eq:Rx}
\end{equation}
Here the space for the indices in (2.5) follows the ``covariant notation'' in physics [1,2]. 
Space rotations in $\mathbb R\times\mathbb R^n$ are given by $R\in{\rm O}(n)$ as
\begin{equation}
    (1\otimes R)(t,\bm x)=(t, R\bm x),\quad x=(t,\bm x)\in\mathbb R\times\mathbb R^n.
\end{equation}
Space-time translations in $\mathbb R\times\mathbb R^n$ are given by $y=(s,\bm y)\in\mathbb R\times\mathbb R^n$ as
\begin{equation}
    T_y\,x=x-y=(t-s,\bm x-\bm y),\quad x=(t,\bm x)\in\mathbb R\times\mathbb R^n.
\label{eq:Tyx}
\end{equation}
Pure Galilei transformations in $\mathbb R\times\mathbb R^n$ are given by $\bm v\in\mathbb R^n$ as
\begin{equation}
G_{\bm v}x=(t,\bm x-t\bm v),\quad x=(t,\bm x)\in\mathbb R\times\mathbb R^n.
\end{equation}
The transformation group generated by $(T_y; y\in\mathbb R\times\mathbb R^n$), $(1\otimes R; R\in{\rm O}(n))$ and $(G_{\bm v}; \bm v\in\mathbb R^n)$ is called the Galilei group in $\mathbb R\times\mathbb R^n$.  
The transformation group generated by $(1\otimes R; R\in{\rm O}(n))$ and $(G_{\bm v}; \bm v\in\mathbb R^n)$ is called the homogeneous Galilei group in $\mathbb R\times\mathbb R^n$.  

The Galilei group is defined to act on functions $u\in{\rm C}^\infty(\mathbb R\times\mathbb R^n;\mathbb C)$ as the corresponding pull-backs:
\begin{align}
    (T_y^*u)(x)&=u(T_yx)=u(x-y)=u(t-s,\bm x-\bm y),\\
    \bigl((1\otimes R)^*u\bigr)(x)&=u((1\otimes R)x)=u(t,R\bm x),\\
    (G_{\bm v}^*u)(x)&=u(G_{\bm v}\bm x)=u(t,\bm x-t\bm v).
\end{align}
We introduce the notion of local gauge as a multiplication of functions of modulus 1 of the form $\exp(i\theta)$ with $\theta\in{\rm C}^\infty(\mathbb R\times\mathbb R^n;\mathbb R)$.

We consider linear partial differential operators in $\mathbb R\times\mathbb R^n$ of order $m$ of the form
\begin{equation}
    L=\sum_{j+|\bm\alpha|\le m}a_{j\bm\alpha}\partial_t^j\bm\partial^{\bm\alpha},
\label{eq:L}
\end{equation}
where $\bm\alpha=(\alpha_1,\ldots,\alpha_n)\in\mathbb Z_{\ge0}^n$ is a multi-index with length $|\bm\alpha|=\alpha_1+\cdots+\alpha_n$, $\bm\partial^{\bm\alpha}=\partial_1^{\alpha_1}\cdots\partial_n^{\alpha_n}$, $\partial_j=\partial/\partial x^j$, $\partial_t=\partial/\partial t$, $a_{j\bm\alpha}\in{\rm C}(\mathbb R\times\mathbb R^n;\mathbb C)$, and $\sum\limits_{j+|\bm\alpha|=m}|a_{j\bm\alpha}|\not=0$.

We say that $L$ is space-time translation invariant if and only if 
\begin{equation}
    T_y^*Lu=LT_y^*u
\end{equation}
for any $y\in\mathbb R\times\mathbb R^n$ and any $u\in{\rm C}^\infty(\mathbb R\times\mathbb R^n;\mathbb C)$.  
We say that $L$ is space-rotation invariant if and only if
\begin{equation}
    (1\otimes R)^*Lu=L(1\otimes R)^*u
\end{equation}
for any $R\in{\rm O}(n)$ and any $u\in{\rm C}^\infty(\mathbb R\times\mathbb R^n;\,\mathbb C)$.  
We say that $L$ is pure Galilei invariant if and only if for any $\bm v\in\mathbb R^n$, there exists $\theta_{\bm v}\in{\rm C}^\infty(\mathbb R\times\mathbb R^n;\mathbb R)$ such that 
\begin{equation}
    e^{i\theta_{\bm v}}G_{\bm v}^*Lu=Le^{i\theta_{\bm v}}G_{\bm v}^*u
\end{equation}
for any $u\in{\rm C}^\infty(\mathbb R\times\mathbb R^n;\mathbb C)$.  
We say that $L$ is Galilei invariant if and only if $L$ is space-time translation invariant, space-rotation invariant, and pure Galilei invariant.  
We say that $L$ is homogeneous Galilei invariant if and only if $L$ is space-rotation invariant and pure Galilei invariant.  Here we remark that an extra degree of freedom $\theta_v$ is necessary because the commutativity between $L$ and pure pullback of translations by constant velocity breakes down.

We now state our main results.
\begin{theorem} \label{t:1}
For any partial differential operator in space-time $\mathbb R\times\mathbb R^n$ of the second order of the form   
\begin{equation}
    L=\sum_{j+|\bm\alpha|\le2}a_{j\bm\alpha}\partial_t^j\bm\partial^{\bm\alpha},
\end{equation}
with $a_{j\bm\alpha}\in{\rm C}(\mathbb R\times\mathbb R^n;\mathbb C)$ and $\sum\limits_{j+|\bm\alpha|=2}|a_{j\bm\alpha}|\not=0$, the following three statements are equivalent.
\begin{itemize}
    \item[(1)] $L$ is Galilei invariant.  
    Namely,
    \begin{itemize}
        \item[(i)] For any $y\in\mathbb R\times\mathbb R^n$ and any $u\in{\rm C}^\infty(\mathbb R\times\mathbb R^n;\mathbb C)$, $T_y^*Lu=LT_y^*u$.
        \item[(ii)] For any $R\in{\rm O}(n)$ and any $u\in{\rm C}^\infty(\mathbb R\times\mathbb R^n;\mathbb C)$, $(1\otimes R)^*u=L(1\otimes R)^*u$.
        \item[(iii)] For any $\bm v\in\mathbb R^n$, there exists $\theta_{\bm v}\in{\rm C}^\infty(\mathbb R\times\mathbb R^n;\mathbb R)$ such that $e^{i\theta_{\bm v}}G_{\bm v}^*Lu=Le^{i\theta_{\bm v}}G_{\bm v}^*u$ for any $u\in{\rm C}^\infty(\mathbb R\times\mathbb R^n;\mathbb C)$.
        \end{itemize}
    \item[(2)] All coefficients of $L$ are constants and $L$ is homogeneous Galilei invariant.  
    Namely,
    \begin{itemize}
    \item[(i)] For any $(j,\bm\alpha)\in\mathbb Z_{\ge0}\times\mathbb Z_{\ge0}^n$ with $j+|\bm\alpha|\le2$ and any $x\in\mathbb R\times\mathbb R^n$, $a_{j\bm\alpha}(x)=a_{j\bm\alpha}(0)$.
    \item[(ii)] For any $R\in{\rm O}(n)$ and any $u\in{\rm C}^\infty(\mathbb R\times\mathbb R^n;\mathbb C)$, $(1\otimes R)^*Lu=L(1\otimes R)^*u$.
    \item[(iii)] For any $\bm v\in\mathbb R^n$, there exists $\theta_{\bm v}\in{\rm C}^\infty(\mathbb R\times\mathbb R^n;\mathbb R)$ such that $e^{i\theta_{\bm v}}G_{\bm v}^*Lu=Le^{i\theta_{\bm v}}G_{\bm v}^*u$ for any $u\in{\rm C}^\infty(\mathbb R\times\mathbb R^n;\mathbb C)$.
    \end{itemize}
    \item[(3)] For any $(j,\bm\alpha)$ with $j+|\bm\alpha|\le2$, $a_{j\bm\alpha}$ is a complex constant and zero except $a_{0\bm0},a_{1\bm0},a_{0,2\bm e_1},\ldots,a_{0,2\bm e_n}$, where $a_{0,2\bm e_1}=\cdots=a_{0,2\bm e_n}\not=0$ hold.
    Let $-\alpha\in\mathbb C\backslash\{0\}$ be the common value: $-\alpha=a_{0,2\bm e_1}=\cdots=a_{0,2\bm e_n}\not=0$.  
    Then ${ia_{1\bm0}/\alpha}\in\mathbb R$, so that if $(\lambda,\beta)\in\mathbb R\times\mathbb C$ are given by $\lambda=-{ia_{1\bm0}\over2\alpha}$ and $\beta=a_{0\bm0}$, $L$ is represented as
    \begin{equation}
        L=\alpha(2i\lambda\partial_t+\triangle)+\beta,
    \end{equation}
    where $\triangle=\nabla^2=\partial_1^2+\cdots\partial_n^2$ is the Laplacian in $\mathbb R^n$.  
    The phase function $\theta_{\bm v}$ in (1)(iii) and (2)(iii)) is given by
    \begin{align}
    \theta_{\bm v}(t,\bm x)&=\theta_{\bm v}(0,\bm0)+\lambda\bm v\cdot\bm x-{\lambda\over2}t|\bm v|^2\quad\hbox{if}\quad\lambda\not=0,\label{eq:thetav}\\
    \theta_{\bm v}(t,\bm x)&=\theta_{\bm v}(t,\bm0)\quad\hbox{if}\quad\lambda=0.
    \end{align}
\end{itemize}
\end{theorem}

\begin{remark} \label{r:1}
    In the case $\lambda=0$ in (3), we have $L=\alpha\triangle+\beta$ and $\theta_{\bm v}(t,\bm x)=\theta_{\bm v}(t,\bm0)$ so that $e^{i\theta_{\bm v}},\,G_{\bm v}^*$, and $L$ are commutative on ${\rm C}^\infty(\mathbb R\times\mathbb R^n;\mathbb C)$. 
\end{remark}

\begin{remark} \label{r:2}
    In the case $\lambda\not=0$ in (3), the constant $\beta$ is essentially removable in the sense that
    \begin{equation}
        \exp\Bigl({\beta\over2i\alpha\lambda}t\Bigr)L\exp\Bigl({-\beta\over2i\alpha\lambda}t\Bigr)u=\alpha(2i\lambda\partial_t+\triangle)u
    \end{equation}
    for all $u\in{\rm C}^\infty(\mathbb R\times\mathbb R^n;\mathbb C)$.  
    If $\beta\in\mathbb R$, the mappings $u\mapsto\exp({\pm\beta\over2i\alpha\lambda})u$ are global gauge transformations and $L$ and $\alpha(2i\lambda\partial_t+\triangle)$ are gauge equivalent.
\end{remark}

\begin{remark} \label{r:3}
    The case $\lambda=1$ is regarded as the standard choice since the essential factor $2i\partial_t+\triangle$ is the time-dependent free Schr\"odinger operator in the scale where the Planck constant $\hbar$ and the mass of the free particle are normalized.  
    Moreover, this choice $\lambda=1$ with another normalization $\theta_{\bm v}(0,\bm0)=0$ yields
    \begin{equation}
        \Bigl(e^{i\theta_{\bm v}}G_{\bm v}^*u\Bigr)(t,\bm x)=\exp\bigl(i(\bm v\cdot\bm x-{t\over2}|\bm v|^2)\bigr)u(t,\bm x-t\bm v).
        \label{eq:Galilei}
    \end{equation}
    The transformation $u\mapsto e^{i\theta_{\bm v}}G_{\bm v}^*u$ in (\ref{eq:Galilei}) is actually called ``the Galilei transformation" in \cite{Carles,Ginibre,LinaresPonce}.
\end{remark}
    
If we assume that $\theta_{\bm v}$ has a specific form as in (\ref{eq:thetav}), we reformulate Theorem~\ref{t:1} to cover higher order partial differential operators as follows. 
\begin{theorem} \label{t:2}
    Let $m\in\mathbb Z_{>0}$ and $\lambda\in\mathbb R\backslash\{0\}$.  
    For any $\bm v\in\mathbb R$, let $\theta_{\bm v}\in{\rm C}^\infty(\mathbb R\times\mathbb R^n;\mathbb R)$ be defined as
    \begin{equation}
        \theta_{\bm v}(t,\bm x)=c+\lambda\bm v\cdot\bm x-{\lambda\over2}t|\bm v|^2,
        \label{eq:cthetav2}
    \end{equation}
    where $c=\theta_{\bm v}(0,\bm0)\in\mathbb C$.  
    Then for any linear partial differential operator in space-time $\mathbb R\times\mathbb R^n$ of order $m$ of the form
    \begin{equation}
        L=\sum_{j+|\bm\alpha|\le m}a_{j\bm\alpha}\partial_t^j\bm\partial^{\bm\alpha}
    \end{equation}
    with $a_{j\bm\alpha}\in{\rm C}(\mathbb R\times\mathbb R^n;\mathbb C)$ and $\sum\limits_{j+|\bm\alpha|=m}|a_{j\bm\alpha}|\not=0$, the following three statements are equivalent.
\begin{itemize}
    \item[(1)] $L$ is Galilei invariant in the following sense.  
    \begin{itemize}
        \item[(i)] For any $y\in\mathbb R\times\mathbb R^n$ and any $u\in{\rm C}^\infty(\mathbb R\times\mathbb R^n;\mathbb C)$, $T_y^*Lu=LT_y^*u$.
        \item[(ii)] For any $R\in{\rm O}(n)$ and any $u\in{\rm C}^\infty(\mathbb R\times\mathbb R^n;\mathbb C)$, $(1\otimes R)^*u=L(1\otimes R)^*u$.
        \item[(iii)] For any $\bm v\in\mathbb R^n$ and any $u\in{\rm C}^\infty(\mathbb R\times\mathbb R^n;\mathbb C)$, $e^{i\theta_{\bm v}}G_{\bm v}^*Lu=Le^{i\theta_{\bm v}}G_{\bm v}^*u$, where $\theta_{\bm v}$ is defined in (\ref{eq:cthetav2}).
        \end{itemize}
    \item[(2)] All coefficients of $L$ are constants and $L$ is homogeneous Galilei invariant in the following sense.  
    \begin{itemize}
    \item[(i)] For any $(j,\bm\alpha)\in\mathbb Z_{\ge0}\times\mathbb Z_{\ge0}^n$ with $j+|\bm\alpha|\le m$ and any $x\in\mathbb R\times\mathbb R^n$, $a_{j\bm\alpha}(x)=a_{j\bm\alpha}(0)$.
    \item[(ii)] For any $R\in{\rm O}(n)$ and any $u\in{\rm C}^\infty(\mathbb R\times\mathbb R^n;\mathbb C)$, $(1\otimes R)^*Lu=L(1\otimes R)^*u$.
        \item[(iii)] For any $\bm v\in\mathbb R^n$ and any $u\in{\rm C}^\infty(\mathbb R\times\mathbb R^n;\mathbb C)$, $e^{i\theta_{\bm v}}G_{\bm v}^*Lu=Le^{i\theta_{\bm v}}G_{\bm v}^*u$, where $\theta_{\bm v}$ is defined in (\ref{eq:cthetav2}).
    \end{itemize}
    \item[(3)] $m$ is an even integer and $L$ is given by a polynomial in $2i\lambda\partial_t+\triangle$ of order $m/2$.  
    Namely, there exists $(a_j;j\in\{0,\ldots,m/2\})\subset\mathbb C$ such that
    \begin{equation*}
        L=\sum_{j=0}^{m/2}a_j(2i\lambda\partial_t+\triangle)^{j}\quad\hbox{\rm and}\quad a_{m/2}\not=0.
    \end{equation*}
\end{itemize}    
\end{theorem}

Theorem~\ref{t:1} shows that the Galilei invariance with local gauge determines the structure of the space of linear partial differential operators of the second order that is exclusively provided with the free Schr\"odinger operator and the formulation of local gauge as well.  
Theorem~\ref{t:2} shows that a natural generalization of Theorem~\ref{t:1} is possible for higher-order operators if we fix the local gauge structure.  
The main results of this paper show that the time-dependent free Schr\"odinger equation is derived from a simple observation and explicit calculations with the action of the Galilei group upon scalar fields up to local gauge, which is independent of the physical convention of substitution of the energy and momentum by the time and space derivatives, respectively.  
Our characterization of the time-dependent free Schr\"odinger equation in terms of the Galilei group with local gauge supports the foundation of Quantum Mechanics on the basis of the invariance naturally arising in Classical Mechanics.

\section{Proof of the Main Results}
\label{s:proof}
In this section, we prove Theorems~\ref{t:1} and \ref{t:2}.  
For that purpose, we introduce some notation.  
For any $\xi=(\tau,\bm\xi)\in\mathbb R\times\mathbb R^n$, we define $e_\xi\in{\rm C}^\infty(\mathbb R\times\mathbb R^n;\mathbb C)$ by
\begin{equation}
    e_\xi(x)=e_\xi(t,\bm x)=\exp(i(\tau t+\bm\xi\cdot\bm x)),\quad x=(t,\bm x)\in\mathbb R\times\mathbb R^n.
    \label{eq:exi}
\end{equation}
Then we see that $e_\xi(0)=1$ and for any $(j,\bm\alpha)\in\mathbb Z_{\ge0}\times\mathbb Z_{\ge0}^n$,
\begin{equation}
    \partial_t^j\bm\partial^{\bm\alpha}e_\xi=(i\tau)^j(i\bm\xi)^{\bm\alpha}e_\xi.
    \label{eq:partialexi}
\end{equation}
For any $x,\xi\in\mathbb R\times\mathbb R^n$, we define
\begin{equation}
    p(x,\xi)=\sum_{j+|\bm\alpha|\le m}a_{j\bm\alpha}(x)(i\tau)^j(i\bm\xi)^{\bm\alpha}.
    \label{eq:pxxi}
\end{equation}
By (\ref{eq:exi}), (\ref{eq:partialexi}), and (\ref{eq:pxxi}), we have
\begin{equation}
    Le_\xi(x)=p(x,\xi)e_\xi(x)
    \label{eq:Lexi}
\end{equation}
for any $x,\xi\in\mathbb R\times\mathbb R^n$ and the original operator $L$ is represented as
\begin{equation}
    L=p(x,-i\partial_t,-i\bm\nabla),
    \label{eq:L=p}
\end{equation}
where $\bm\nabla=\bm\partial=(\partial_1,\ldots,\partial_n)$.  
We use the equalities (\ref{eq:Lexi}) and (\ref{eq:L=p}) in the subsequent argument.   

\medskip
\noindent\underline{\it Proof of Theorem~\ref{t:1}\/}

\medskip
\noindent$(1)\Rightarrow(2)$: It suffices to prove (1)(i)$\Rightarrow$(2)(i).  
The translation invariance in (1)(i) with $u=e_\xi$ implies
\begin{equation}
    T_y^*Le_\xi=LT_y^*e_\xi
\label{eq:TLexi}
\end{equation}
for any $y=(s,\bm y)\in\mathbb R\times\mathbb R^n$. 
The LHS of (\ref{eq:TLexi}) at $x=(t,\bm x)$ is evaluated as
\begin{equation}
    (T_y^*Le_\xi)(x)=(Le_\xi)(x-y)=p(x-y,\xi)e_\xi(x-y),
    \label{eq:TLexi2}
\end{equation}
where we have used (\ref{eq:L=p}), while the RHS of (\ref{eq:TLexi}) at $x=(t,\bm x)$ is evaluated as
\begin{align}
    (LT_y^*e_\xi)(x)&=\sum_{j+|\bm\alpha|\le2}a_{j\bm\alpha}(x)(\partial_t^j\bm\partial^{\bm\alpha}T_y^*e_\xi)(x)=\sum_{j+|\bm\alpha|\le2}a_{j\bm\alpha}(x)(T_y^*\partial_t^j\bm\partial^{\bm\alpha}e_\xi)(x)\nonumber\\
    &=\sum_{j+|\bm\alpha|\le2}a_{j\bm\alpha}(x)(i\tau)^j(i\bm\xi)^{\bm\alpha}e_\xi(x-y)=p(x,\xi)e_\xi(x-y),
\label{eq:LTexi}
\end{align}
where we have used (\ref{eq:partialexi}) and (\ref{eq:pxxi}).  
By (\ref{eq:TLexi}), (\ref{eq:TLexi2}), and (\ref{eq:LTexi}), with $y=x$, we have
\begin{equation}
    p(0,\xi)=p(x,\xi)
\label{eq:p0xi}
\end{equation}
for any $\xi=(\tau,\bm\xi)\in\mathbb R\times\mathbb R^n$, since $e_\xi(0)=1$.  
It follows from (\ref{eq:p0xi}) that
\begin{equation}
    a_{j\bm\alpha}(x)=a_{j\bm\alpha}(0)
\end{equation}
for any $x\in\mathbb R\times\mathbb R^n$ and any $(j,\bm\alpha)\in\mathbb Z_{\ge0}\times\mathbb Z_{\ge0}^n$, which is precisely the statement in (2)(i).

\medskip
\noindent$(2)\Rightarrow(3)$: By (2)(i), we denote by $p(\xi)=p(0,\xi)$ the following polynomial in $\xi=(\tau,\bm\xi)$ of the second order
\begin{equation}
    p(\xi)=\sum_{j+|\bm\alpha|\le2}a_{j\bm\alpha}(i\tau)^j(i\bm\xi)^{\bm\alpha}
    \label{eq:pxi}
\end{equation}
with $(a_{j\bm\alpha};j+|\bm\alpha|\le2)\subset\mathbb C$.  
For $j\in\{0,1,2\}$, we define
\begin{equation}
    p_j(\bm\xi)=\sum_{|\bm\alpha|\le2-j}a_{j\bm\alpha}(0)(i\bm\xi)^{\bm\alpha},
\end{equation}
so that $p$ is represented as
\begin{equation}
    p(\xi)=\sum_{j=0}^2p_j(\bm\xi)(i\tau)^j=p_0(\bm\xi)+ip_1(\bm\xi)\tau-p_2(\bm\xi)\tau^2.
\label{eq:pxi1}
\end{equation}
We prove that $p_j$ is invariant under the space rotations:
\begin{equation}
    R^*p_j=p_j
\label{eq:Rpj}
\end{equation}
for any $R\in{\rm O}(n)$.  
For that purpose we calculate
\begin{align}
    \bigl((1\otimes R)^*e_\xi\bigr)(x)&=e_\xi(t,R\bm x)=\exp(i(\tau t+\bm\xi\cdot R\bm x))\nonumber\\
    &=\exp(i(\tau t+\,{}^t\!R\bm\xi\cdot\bm x))=e_{(\tau,\,{}^t\!R\bm\xi)}(x).
    \label{eq:Rexi}
\end{align}
By (\ref{eq:partialexi}) and (\ref{eq:Rexi}), we obtain
\begin{equation}
    \partial_t^j\bm\partial^{\bm\alpha}(1\otimes R)^*e_\xi=(i\tau)^j(i\,{}^t\!R\bm\xi)^{\bm\alpha}e_{(\tau,{}^tR\bm\xi)}
\end{equation}
and therefore
\begin{equation}
    L(1\otimes R)^*e_\xi=\sum_{j+|\bm\alpha|\le2}a_{j\bm\alpha}(i\tau)^j(i\,{}^t\!R\bm\xi)^{\bm\alpha}e_{(\tau,\,{}^t\!R\bm\xi)}=p(\tau,{}^t\!R\bm\xi)e_{(\tau,\,{}^t\!R\bm\xi)},
    \label{eq:LRexi}
\end{equation}
which is equal to $(1\otimes R)^*Le_\xi$ by (2)(ii), given by
\begin{equation}
    (1\otimes R)^*Le_\xi=p(\xi)(1\otimes R)^*e_\xi=p(\xi)e_{(\tau,\,^t\!R\bm\xi)},
    \label{eq:RLexi}
\end{equation}
where we have used (\ref{eq:Lexi}), (\ref{eq:pxi}), and (\ref{eq:Rexi}).  
The equality that (\ref{eq:LRexi}) and (\ref{eq:RLexi}) form implies, at $x=0\in\mathbb R\times\mathbb R^n$,
\begin{equation}
    \sum_{j=0}^2p_j({}^t\!R\bm\xi)(i\tau)^j=p(\tau,{}^t\!R\bm\xi)=p(\tau,\bm\xi)=\sum_{j=0}^2p_j(\bm\xi)(i\tau)^j,
    \label{eq:p=p}
\end{equation}
where we have used (\ref{eq:pxi1}).  
Since (\ref{eq:p=p}) holds for any $\xi=(\tau,\bm\xi)\in\mathbb R\times\mathbb R^n$ and any $R\in{\rm O}(n)$, we conclude that
\begin{equation}
    (\,{}^t\!R)^*p_j=p_j
    \label{eq:Rpjpj}
\end{equation}
holds for any $R\in{\rm O}(n)$ and any $j\in\{1,2,3\}$.  
By replacing $R\mapsto{}^t\!R$, (\ref{eq:Rpjpj}) is equivalent to (\ref{eq:Rpj}), as required.

Since $p_j$ is invariant under space rotations as in (\ref{eq:Rpj}), it is written as a polynomial in $|\bm\xi|^2$ as
\begin{align}
    p_0(\bm\xi)&=b_{00}+b_{01}|\bm\xi|^2,\label{eq:p0}\\
    p_1(\bm\xi)&=b_{10},\label{eq:p1}\\
    p_2(\bm\xi)&=b_{20}\label{eq:p2}
\end{align}
with $b_{00},b_{01},b_{10},b_{20}\in\mathbb C$ (see \cite{NakazatoOzawa,Nomura,Shimakura}), as the orders of $p_0,p_1,p_2$ are $2,0,0$, respectively.  
By (\ref{eq:pxi1}), (\ref{eq:p0}), (\ref{eq:p1}), and (\ref{eq:p2}), $p$ is represented as
\begin{equation}
p(\tau,\bm\xi)=(b_{00}+b_{01}|\bm\xi|^2)+ib_{10}\tau-b_{20}\tau^2.
\label{eq:ptauxi}
\end{equation}
By the definition of $p$ in (\ref{eq:pxi}), we see from (\ref{eq:ptauxi}) that $b_{00}=a_{0\bm0}$, $b_{01}=a_{0,2\bm e_1}=\cdots=a_{0,2\bm e_n}$, $b_{10}=a_{1\bm0}$, and $b_{20}=a_{2\bm0}$.  
We now define $\alpha$ and $\beta$ by 
\begin{equation}
    \alpha=-b_{01}=-a_{0,2\bm e_1}=\cdots=-a_{0,2\bm e_n}\quad\hbox{\rm and}\quad\beta=a_{0\bm0}.
    \label{eq:alphabeta}
\end{equation}
Then, by (\ref{eq:ptauxi}) and (\ref{eq:alphabeta}), $p$ is given by
\begin{equation}
    p(\tau,\bm\xi)=\beta-\alpha|\bm\xi|^2+ia_{1\bm0}\tau-a_{2\bm0}\tau^2.
    \label{eq:ptauxi2}
\end{equation}
By (\ref{eq:L=p}) and (\ref{eq:ptauxi2}), $L$ is given by
\begin{equation}
    L=p(-i\partial_t,-i\bm\nabla)=\beta+\alpha\triangle+a_{1\bm0}\partial_t+a_{2\bm0}\partial_t^2.
    \label{eq:L2}
\end{equation}
With the representation (\ref{eq:L2}), we calculate both sides of the equality in (2)(iii) with $u=e_\xi$ to obtain an explicit formula for $L$ as well as $\theta_{\bm v}$.  
We carry out differentiation of the function $e^{i\theta_{\bm v}}G_{\bm v}^*e_\xi$ given by
\begin{equation}
    \bigl(e^{i\theta_{\bm v}}G_{\bm v}^*e_\xi\bigr)(x)=\exp\bigl(i\theta_{\bm v}(t,\bm x)+i\tau t+i\bm\xi\cdot(\bm x-t\bm v)\bigr)
\end{equation}
with any $\theta_{\bm v}\in{\rm C}^\infty(\mathbb R\times\mathbb R^n;\mathbb R)$ to obtain
\begin{align}
    \bm\nabla\bigl(e^{i\theta_{\bm v}}G_{\bm v}^*e_\xi\bigr)&=\bigl(e^{i\theta_{\bm v}}G_{\bm v}^*e_\xi\bigr)(i\bm\nabla\theta_{\bm v}+i\bm\xi),\\
    \triangle\bigl(e^{i\theta_{\bm v}}G_{\bm v}^*e_\xi\bigr)&=\bigl(e^{i\theta_{\bm v}}G_{\bm v}^*e_\xi\bigr)(i\triangle\theta_{\bm v}-|\nabla\theta_{\bm v}|^2-2\bm\xi\cdot\bm\nabla\theta_{\bm v}-|\bm\xi|^2),\\
    \partial_t\bigl(e^{i\theta_{\bm v}}G_{\bm v}^*e_\xi\bigr)&=\bigl(e^{i\theta_{\bm v}}G_{\bm v}^*e_\xi\bigr)(i\partial_t\theta_{\bm v}+i\tau-i\bm\xi\cdot\bm v),\\
    \partial_t^2\bigl(e^{i\theta_{\bm v}}G_{\bm v}^*e_\xi\bigr)&=\bigl(e^{i\theta_{\bm v}}G_{\bm v}^*e_\xi\bigr)(i\partial_t^2\theta_{\bm v}-(\partial_t\theta_{\bm v})^2-\tau^2-(\bm\xi\cdot\bm v)^2\nonumber\\
    &\qquad\qquad\qquad-2\tau\partial_t\theta_{\bm v}+2\tau\bm\xi\cdot\bm v+2\partial_t\theta_{\bm v}\bm\xi\cdot\bm v),
\end{align}
which together with (\ref{eq:L2}) yield
\begin{align}
    L\bigl(e^{i\theta_{\bm v}}G_{\bm v}^*e_\xi\bigr)=\bigl(e^{i\theta_{\bm v}}G_{\bm v}^*e_\xi\bigr)
    \bigl(&\beta+\alpha(i\triangle\theta_{\bm v}-|\nabla\theta_{\bm v}|^2-2\bm\xi\cdot\bm\nabla\theta_{\bm v}-|\bm\xi|^2)\nonumber\\
    &+ia_{1\bm0}(\partial_t\theta_{\bm v}+\tau-\bm\xi\cdot\bm v)\nonumber\\
    &+a_{2\bm0}(i\partial_t^2\theta_{\bm v}-(\partial_t\theta_{\bm v})^2-\tau^2-(\bm\xi\cdot\bm v)^2\nonumber\\
    &\qquad\quad-2\tau\partial_t\theta_{\bm v}+2\tau\bm\xi\cdot\bm v+2\partial_t\theta_{\bm v}\bm\xi\cdot\bm v)\bigr).
    \label{eq:LGe}
\end{align}
By (2)(iii), (\ref{eq:LGe}) is equal to 
\begin{equation}
    e^{i\theta_{\bm v}}G_{\bm v}^*Le_\xi=\bigl(e^{i\theta_{\bm v}}G_{\bm v}^*e_\xi\bigr)(\beta-\alpha|\xi|^2+ia_{1\bm0}\tau-a_{2\bm0}\tau^2).
    \label{eq:GLe}
\end{equation}
It follows from (\ref{eq:LGe}) and (\ref{eq:GLe}) that the equality
\begin{align}
    &\alpha(i\triangle\theta_{\bm v}-|\nabla\theta_{\bm v}|^2-2\bm\xi\cdot\bm\nabla\theta_{\bm v})+ia_{1\bm0}(\partial_t\theta_{\bm v}-\bm\xi\cdot\bm v)\nonumber\\
    &+a_{2\bm0}(i\partial_t^2\theta_{\bm v}-(\partial_t\theta_{\bm v})^2-(\bm\xi\cdot\bm v)^2-2\tau\partial_t\theta_{\bm v}+2\tau\bm\xi\cdot\bm v+2\partial_t\theta_{\bm v}\bm\xi\cdot\bm v)=0
    \label{eq:cond1}
\end{align}
holds for any $\bm v,\bm\xi\in\mathbb R^n$ and any $t\in\mathbb R$.
By considering the quadratic contribution of $\bm\xi$ in (\ref{eq:cond1}), we conclude $a_{2\bm0}=0$.  
Then (\ref{eq:cond1}) is rewritten as
\begin{equation}
    \alpha(i\triangle\theta_{\bm v}-|\nabla\theta_{\bm v}|^2)+ia_{1\bm0}\partial_t\theta_{\bm v}-\bm\xi\cdot(2\alpha\bm\nabla\theta_{\bm v}+ia_{1\bm0}\bm v)=0.
    \label{eq:cond2}
\end{equation}
It follows from (\ref{eq:cond2}) that $\theta_{\bm v}$ must satify
\begin{align}
    2\alpha\bm\nabla\theta_{\bm v}+ia_{1\bm0}\bm v&=0,\label{eq:cond3a}\\
    \alpha(i\triangle\theta_{\bm v}-|\nabla\theta_{\bm v}|^2)+ia_{1\bm0}\partial_t\theta_{\bm v}&=0,
    \label{eq:cond3b}
\end{align}
since (\ref{eq:cond2}) holds for any $\bm\xi\in\mathbb R^n$.  
Since $a_{2\bm0}=0$, (\ref{eq:ptauxi2}) and (\ref{eq:L2}) are rewritten as
\begin{align}
    p(\tau,\bm\xi)&=\beta-\alpha|\bm\xi|^2+ia_{1\bm0}\tau,\label{eq:ptauxi3}\\
    L&=\beta+\alpha\triangle+a_{1\bm0}\partial_t,\label{eq:L3}
\end{align}
respectively, which in turn ensure that $\alpha\not=0$ because $L$ is of the second order.  
This implies that (\ref{eq:cond3a}) is equivalent to
\begin{equation}
    \bm\nabla\bigl(\theta_{\bm v}+{ia_{1\bm0}\over2\alpha}\bm v\cdot\bm x\bigr)=0,
\end{equation}
which immediately proves that $\theta_{\bm v}$ is given by
\begin{equation}
    \theta_{\bm v}(t,\bm x)=\theta_{\bm v}(t,\bm0)-{ia_{1\bm0}\over2\alpha}\bm v\cdot\bm x.
    \label{eq:thetav1}
\end{equation}
The assumption that $\theta_{\bm v}$ is real-valued implies that the coefficient of $\bm v\cdot\bm x$ on the RHS of (\ref{eq:thetav1}) must be real and we introduce 
\begin{equation}
    \lambda=-{ia_{1\bm0}\over2\alpha}\in\mathbb R.
\end{equation}
This shows that
\begin{align}
    \theta_{\bm v}(t,\bm x)&=\theta_{\bm v}(t,\bm0)+\lambda\bm v\cdot\bm x,\label{eq:thetav2}\\
    \bm\nabla\theta_{\bm v}(t,\bm x)&=\lambda\bm v,\label{eq:nablatheta}\\
    \triangle\theta_{\bm v}(t,\bm x)&=0.\label{eq:laplasethetav}
\end{align}
Then, the equation (\ref{eq:cond3b}) takes the form
\begin{equation}
    |\bm\nabla\theta_{\bm v}|^2+2\lambda\partial_t\theta_{\bm v}=\lambda^2|\bm v|^2+2\lambda\partial_t\theta_{\bm v}(t,\bm0)=0,
\end{equation}
which shows that $\theta_{\bm v}(t,\bm0)$ is given by
\begin{equation}
    \theta_{\bm v}(t,\bm0)=\theta_{\bm v}(0,\bm0)-{\lambda\over2}t|\bm v|^2\quad\hbox{\rm if}\quad\lambda\not=0.
\end{equation}
This implies that $\theta_{\bm v}$ is given by
\begin{align}
    \theta_{\bm v}(t,\bm x)&=\theta_{\bm v}(0,\bm0)+\lambda\bm v\cdot\bm x-{\lambda\over2}t|\bm v|^2\quad\hbox{\rm if}\quad\lambda\not=0,\label{eq:thetav2a}\\
    \theta_{\bm v}(t,\bm x)&=\theta_{\bm v}(t,\bm0)\quad\hbox{\rm if}\quad\lambda=0.\label{eq:thetav2b}
\end{align}
Moreover, $p$ and $L$ are given by
\begin{align}
    p(\tau,\bm x)&=\beta+\alpha(-2\lambda\tau-|\bm\xi|^2),\\
    L&=\beta+\alpha(2i\lambda\partial_t+\triangle),
\end{align}
respectively.  
This completes the proof of $(2)\Rightarrow(3)$.

\medskip
\noindent$(3)\Rightarrow(1)$: Since $(2)\Rightarrow(1)$, it suffices to prove $(3)\Rightarrow(2)$.  
By the commutativity between $\triangle$ and $R^*$, the statement (2)(ii) follows immediately.  
Therefore, we only have to show (2)(iii).  
Let $\theta\in{\rm C}^\infty(\mathbb R\times\mathbb R^n;\mathbb R)$ and $u\in{\rm C}^\infty(\mathbb R\times\mathbb R^n;\mathbb C)$ and differentiate the function $e^{i\theta}G_{\bm v}^*u$ given by
\begin{equation}
    \bigl(e^{i\theta}G_{\bm v}^*u\bigr)(t,\bm x)=e^{i\theta(t,\bm x)}u(t,\bm x-t\bm v)
\end{equation}
to obtain
\begin{align}
    \bm\nabla\bigl(e^{i\theta}G_{\bm v}^*u\bigr)&=e^{i\theta}(i\bm\nabla\theta G_{\bm v}^*u+G_{\bm v}^*\bm\nabla u),\\
    \triangle\bigl(e^{i\theta}G_{\bm v}^*u\bigr)&=e^{i\theta}\bigl((i\triangle\theta-|\bm\nabla\theta|^2)G_{\bm v}^*u+2i\bm\nabla\theta\cdot G_{\bm v}^*\bm\nabla u+G_{\bm v}^*\triangle u\bigr),\\
    \partial_t\bigl(e^{i\theta}G_{\bm v}^*u\bigr)&=(i\partial_t\theta G_{\bm v}^*u+G_{\bm v}^*\partial_t u-\bm v\cdot G_{\bm v}^*\bm\nabla u)
\end{align}
so that the action of $L=\alpha(2i\lambda\partial_t+\triangle)+\beta$ on $e^{i\theta}G_{\bm v}^*u$ is given by
\begin{align}
    &L\bigl(e^{i\theta}G_{\bm v}^*u\bigr)\nonumber\\
    &=e^{i\theta}G_{\bm v}^*Lu+\alpha e^{i\theta}(-2\lambda\partial_t\theta+i\triangle\theta-|\bm\nabla\theta|^2)G_{\bm v}^*u+2i\alpha e^{i\theta}(\bm\nabla\theta-\lambda\bm v)\cdot G_{\bm v}^*\bm\nabla u.
\end{align}
For $\lambda\in\mathbb R\backslash\{0\}$, we define $\theta$ as $\theta(t,\bm x)=\theta(0,\bm0)+\lambda\bm v\cdot\bm x-{\lambda\over2}t|\bm v|^2$ to obtain
\begin{align}
    -2\lambda\partial_t\theta+i\triangle\theta-|\bm\nabla\theta|^2&=-2\lambda(-{\lambda\over2}|\bm v|^2)-|\lambda\bm v|^2=0,\\
    \bm\nabla\theta-\lambda\bm v&=\lambda\bm v-\lambda \bm v=0.
\end{align}
For $\lambda=0$, we define $\theta$ as $\theta(t,\bm x)=\theta(t,\bm0)$ to obtain
\begin{align}
    -2\lambda\partial_t\theta+i\triangle\theta-|\bm\nabla\theta|^2&=i\triangle\theta-|\bm\nabla\theta|^2=0,\\
    \bm\nabla\theta-\lambda\bm v&=-\lambda \bm v=0.
\end{align}
In both cases, we have $L(e^{i\theta}G_{\bm v}^*u)=e^{i\theta}G_{\bm v}^*Lu$, as required.

\medskip
\noindent\underline{\it Proof of Theorem~\ref{t:2}\/}

\smallskip\noindent 
It suffices to prove that $(2)\Rightarrow(3)$ since other statements are proved in the same way as in the proof of Theorem~\ref{t:1}.  
Let $\theta_{\bm v}$ be as in (\ref{eq:cthetav2}). 
Then $e^{i\theta_{\bm v}}G_{\bm v}^*e_\xi$ is given by
\begin{align}
    \bigl(e^{i\theta_{\bm v}}G_{\bm v}^*e_\xi\bigr)(x)&=\exp\bigl(i(c+\lambda\bm v\cdot\bm x-{\lambda\over2}t|\bm v|^2)\bigr)\exp\bigl(i(\tau t+\bm\xi\cdot(\bm x-t\bm v))\bigr)\nonumber\\
    &=e^{ic}\exp\bigl(i((\tau-\bm\xi\cdot\bm v-{\lambda\over2}|\bm v|^2)t+(\bm\xi+\lambda\bm v)\cdot\bm x\bigr)\nonumber\\
    &=e^{ic}e_{(\tau-\bm\xi\cdot\bm v-{\lambda\over2}|\bm v|^2,\;\bm\xi+\lambda\bm v)}(x).
\end{align}
This leads to
\begin{equation}
    Le^{i\theta_{\bm v}}G_{\bm v}^*e_\xi=e^{ic}\sum_{j+|\bm\alpha|\le m}a_{j\bm\alpha}\bigl(i(\tau-\bm\xi\cdot\bm v-{\lambda\over2}|\bm v|^2)\bigr)^j(\bm\xi+\lambda\bm v)^{\bm\alpha}e_{(\tau-\bm\xi\cdot\bm v-{\lambda\over2}|\bm v|^2,\;\bm\xi+\lambda\bm v)},
    \label{eq:LeGexi}
\end{equation}
where $a_{j\bm\alpha}\in\mathbb C$ by (2)(i).  
By (2)(iii), the LHS of (\ref{eq:LeGexi}) is equal to 
\begin{equation}
    e^{i\theta_{\bm v}}G_{\bm v}^*Le_\xi=p(\xi)e^{i\theta_{\bm v}}G_{\bm v}^*e_\xi=e^{ic}p(\xi)e_{(\tau-\bm\xi\cdot\bm v-{\lambda\over2}|\bm v|^2,\;\bm\xi+\lambda\bm v)}.
    \label{eq:eGLexi}
\end{equation}
The resulting equality between (\ref{eq:LeGexi}) and (\ref{eq:eGLexi}) evaluated at $x=0\in\mathbb R\times\mathbb R^n$ becomes
\begin{align}
    p(\xi)&=\sum_{j=0}^mp_j(\bm\xi)(i\tau)^j=p(\tau-\bm\xi\cdot\bm v-{\lambda\over2}|\bm v|^2,\;\bm\xi+\lambda\bm v)\nonumber\\
    &=\sum_{j=0}^mp_j(\bm\xi+\lambda\bm v)\bigl(i(\tau-\bm\xi\cdot\bm v-{\lambda\over2}|\bm v|^2)\bigr)^j
    \label{eq:pp}
\end{align}
for any $\xi=(\tau,\bm\xi)\in\mathbb R\times\mathbb R^n$ and any $\bm v\in\mathbb R^n$.  
We already know from (2)(ii) that $p_j$ is invariant under space rotations and represented as a polynomial of $|\bm\xi|^2$ of the form
\begin{equation}
    p_j(\bm\xi)=\sum_{2k\le m-j}b_{jk}|\bm\xi|^{2k}
    \label{eq:pj}
\end{equation}
with $b_{jk}\in\mathbb C$.  
By (\ref{eq:pp}) and (\ref{eq:pj}), we have the equality
\begin{equation}
    \sum_{j=0}^m\sum_{2k\le m-j}b_{jk}|\bm\xi|^{2k}(i\tau)^j=\sum_{j=0}^m\sum_{2k\le m-j}b_{jk}|\bm\xi+\lambda\bm v|^{2k}\bigl(i(\tau-\bm\xi\cdot\bm v-{\lambda\over2}|\bm v|^2)\bigr)^j
    \label{eq:equality}
\end{equation}
for any $\xi=(\tau,\bm\xi)\in\mathbb R\times\mathbb R^n$ and $\bm v\in\mathbb R^n$.  
If we choose $\bm v=-{1\over\lambda}\bm\xi$, we have $\bm\xi+\lambda\bm v=0$ and $-\bm\xi\cdot\bm v-{\lambda\over2}|\bm v|^2={1\over2\lambda}|\bm\xi|^2$ and (\ref{eq:equality}) becomes
\begin{equation}
    \sum_{j=0}^m\sum_{2k\le m-j}b_{jk}|\bm\xi|^{2k}(i\tau)^j=\sum_{j=0}^m b_{j0}\bigl(i(\tau+{1\over2\lambda}|\bm\xi|^2)\bigr)^j.
    \label{eq:equality1}
\end{equation}
Here we recall that in the context of Theorem 2, $\lambda \ne 0$.
Since the order of the polynomial $p$ is $m$, the order of the RHS of (\ref{eq:equality1}) must satisfy $2j\le m$, which means that $b_{j0}=0$ if $j>m/2$, implying that $m$ is even and $p$ is given by the polynomial of $2\lambda\tau+|\bm\xi|^2$ of order $m/2$.  
Since $L$ is written as $L=p(-i\partial_t,-i\bm\nabla)$, the statement (3) holds.

\section*{Acknowledgments}
\noindent We wish to thank the referees for important remarks and suggestions.

\noindent H.N. is partially supported by JSPS KAKENHI Grant No.~23K03268.
T.O. is partially supported by JSPS KAKENHI Grant No.~24H00024.

\section*{Declarations}

\noindent{\bf H. Nakazato}: Writing-original draft (support); formal analysis (support); writing-review and editing (equal)
{\bf T. Ozawa}: Conceptualization (lead); writing-original draft (lead); formal analysis (lead); writing-review and editing (equal).

\bigskip\noindent{\bf Conflict of Interest} 

\noindent The authors declare that they have no known competing financial interests or personal relationships that could have appeared to influence the work reported in this paper.

\bigskip\noindent{\bf Data Availability}

\noindent Data sharing is not applicable to this article as no new data were created or analyzed in this study.

\end{document}